\documentclass[12pt
,notitlepage
,superscriptaddress,
showkeys
]{revtex4}
\usepackage{graphicx}
\usepackage{subfigure}
\usepackage{amsmath}
\usepackage{comment}

\begin{document}

\title{Are Room Temperature Ionic Liquids Dilute Electrolytes?}
\keywords{Ion Pairs, Associating Liquids, Electrolyte Solutions, McMilan-Mayer model, Poisson-Boltzmann theory}

\author{Alpha A Lee }
\affiliation{Mathematical Institute, University of Oxford, Oxford OX2 6GG, United Kingdom}

\author{Dominic Vella}
\affiliation{Mathematical Institute, University of Oxford, Oxford OX2 6GG, United Kingdom}

\author{Susan Perkin}
\affiliation{Department of Chemistry, University of Oxford, Oxford, OX1 3QZ, United Kingdom}

\author{Alain Goriely}
\affiliation{Mathematical Institute, University of Oxford, Oxford OX2 6GG, United Kingdom}

\begin{abstract}
An important question in understanding the structure of ionic liquids is whether ions are truly ``free'' and mobile  which would correspond to a concentrated ionic melt, or are rather ``bound'' in ion pairs, that is a liquid of ion pairs with a small concentration of free ions. Recent surface force balance experiments from different groups have given conflicting answers to this question. We propose a simple model for the thermodynamics and kinetics of ion pairing in ionic liquids. Our model takes into account screened ion-ion, dipole-dipole and dipole-ion interactions in the mean field limit. The results of this model suggest that almost two thirds of the ions are free at any instant, and ion pairs have a short lifetime comparable to the characteristic timescale for diffusion. These results suggest that there is no particular thermodynamic or kinetic preference for ions residing in pairs. We therefore conclude that ionic liquids are concentrated, rather than dilute, electrolytes. 
\end{abstract}

\makeatother
\maketitle

%\section{Introduction}

Room temperature ionic liquids are salts that are in the liquid state under ambient conditions. They are important in diverse chemical and physical applications \cite{fedorov2014ionic}, from solvents in organic synthesis \cite{welton1999room,hallett2011room} to applications in field effect transistors \cite{fujimoto2013electric}. Naively, an ionic liquid appears to be a fluid of free ions. However, the presence of strong and long-ranged Coulomb interactions complicates the picture. Recent surface force balance experiments on ionic liquids confined between charged surfaces led to the conclusion \cite{gebbie2013ionic} that ionic liquids behave as \emph{dilute} electrolytes: the majority of ions in solution are bound as cation-anion pairs that behave effectively as dipoles. The authors concluded that these dipoles behave as a solvent containing a relatively low concentration of truly free ions. This conclusion followed from fitting  measured double layer forces to the linear Debye-H\"{u}ckel theory: a very large Debye screening length, of $O(10 \mathrm{nm})$, was found from this fit, corresponding to a very low concentration of free charges. 

This description of ionic liquids triggered an intense discussion in the literature \cite{perkin2013stern,gebbie2013reply}. A key argument for the dilute electrolyte picture is the apparently large dissociation constant. The equilibrium constant for dissociation was estimated as \cite{gebbie2013ionic}
\begin{equation}
K = \exp\left(- \frac{\Delta E_d}{ \epsilon k_B T} \right) 
\label{gebbie_K}
\end{equation}
with dissociation energy of the ion pair $\Delta E_d = 315.26 \; \mathrm{k J} \mathrm{mol}^{-1} $ (taken from quantum electronic structure calculations of an ion pair in vacuum \cite{hunt2007structure}), and $\epsilon = 11.6$, the low frequency dielectric constant (measured using dielectric spectroscopy) \cite{weingartner2008understanding}.
%However, several observations must be made: The value of $E_d$ is the theoretical gas phase dissociation energy. The solution phase dissociation energy is likely to be much smaller due to electrostatic screening by other free ions \cite{lynden2010screening}. Therefore, a dielectric constant alone, as measured from dielectric spectroscopy, is unlikely to be able to account for the electrostatic interaction between ions. \cite{schroder2008collective,lynden2010screening} In fact, some theories and simulations \cite{lynden2007can, lynden2007does} show that taking the low frequency dielectric constant $\epsilon = \infty$ gives good agreement with electron transfer kinetics in ionic liquids! The unbounded dielectric constant physically reminds us of the importance of \emph{free} charges in the medium. Physically, Equation (\ref{gebbie_K}) only holds when the  concentration of free charges is low such as the typical interaction between of an ion pair to another free charge in the medium is less than the thermal energy. It is thus not surprising that that \emph{a priori} assumption lead to the conclusion of low concentration of charges.
However, it is important to recall that in arriving at this estimate, several assumptions have been made. First, the value of $\Delta E_d$ used is the theoretical gas phase dissociation energy, which is likely to be an overestimate because it neglects the electrostatic screening by other free ions \cite{lynden2010screening}. Further, dividing the interaction energy by the dielectric constant, as measured from dielectric spectroscopy, is unlikely to account correctly for the electrostatic interaction between ions \cite{schroder2008collective,lynden2010screening}.  Therefore, Equation (\ref{gebbie_K}) only holds when the concentration of free charges is low, since it implicitly assumes that the typical interaction energy between a bound ion and a neighbouring free charge  is significantly less than the thermal energy \cite{fowler1949statistical}.  
%Thus it is perhaps not surprising that this \emph{a priori} assumption led to the conclusion that ionic liquids behave as dilute electrolytes.

In this Letter, we attempt to remove some of these implicit assumptions. The key questions that we would like to answer are: Are ion pairs abundant in solution? If so, are they long-lived species, or merely transient intermediates? To estimate the abundance and the lifetime of ion pairs in ionic liquids, we develop a simple theory of ion pairing that accounts for both screened electrostatic interactions and a dielectric constant that is self-consistently calculated with the concentration of ion pair dipoles. 

We note that the notions of association equilibrium and ``ion-pairs'' are somewhat artificial: in reality, the liquid is a sea of ions interacting via a Coulomb potential. Nonetheless the idea of ion pairing is an extremely useful concept as it provides a direct chemical and physical analogy with dilute/concentrated electrolytes. 
%A similar concept of association equilibrium has been used successfully to model criticality in a Coulomb fluid. \cite{levin1996criticality}

\begin{figure}
\includegraphics[scale=0.45]{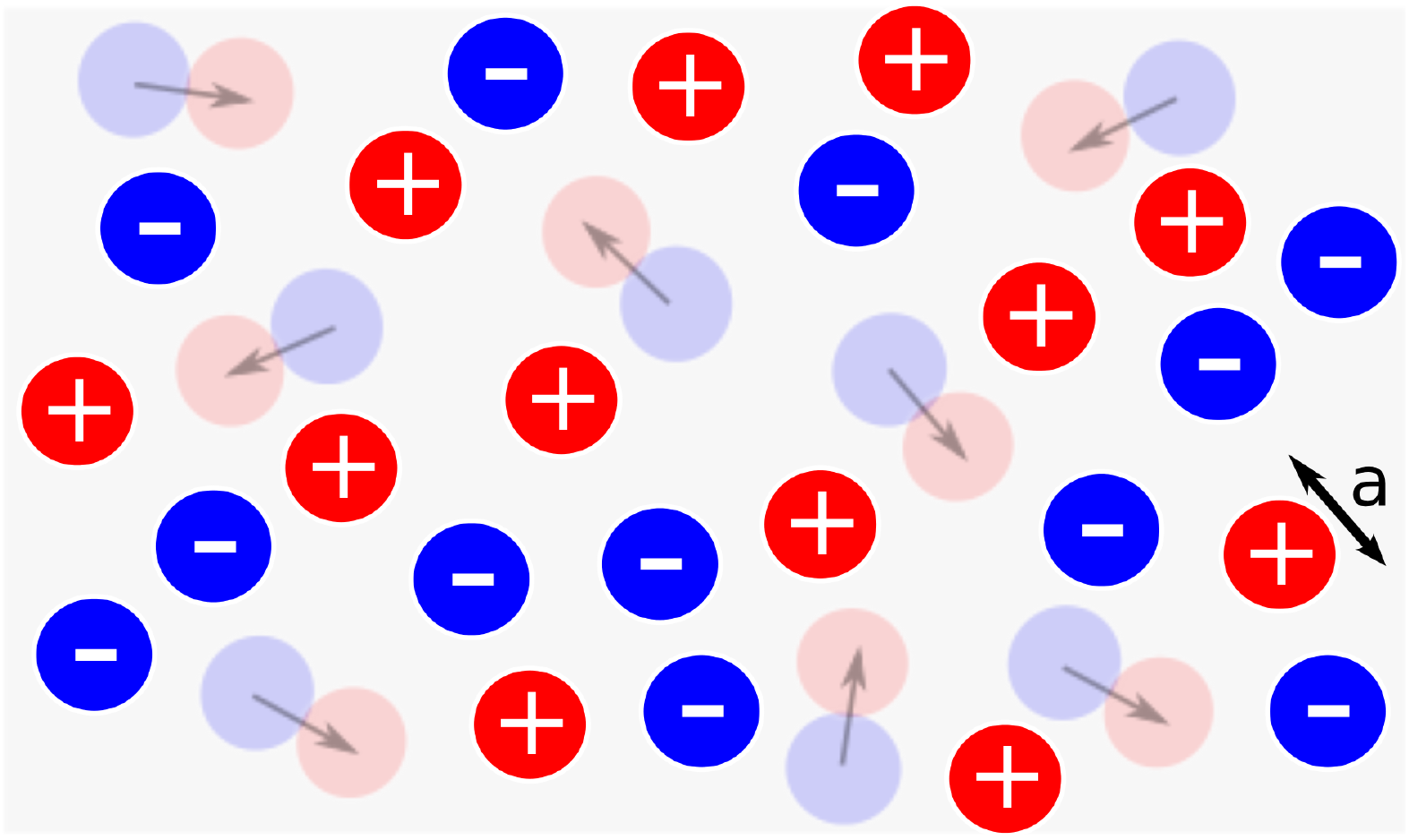}
\caption{An ionic liquid is modelled as a mixture of free ions (coloured spheres) and bounded ion pair dipoles (dumbbells). The free ions provide a characteristic Debye screening length $\kappa^{-1}$ while the dipoles provide a background dielectric constant $\epsilon$.}
\label{cartoon}
\end{figure}

%\section{Thermodynamics of Ion Association}
The picture we suggest is as follows: in an ionic liquid, ions interact with one another screened by other free ions and a background dielectric medium that consists of dipolar ion pairs.  At the same time, dipoles (ion pairs) interact with one another, an interaction that is screened by the presence of free ions and other dipoles (see Fig. \ref{cartoon} for a schematic of the system). Related ideas of ion association have been employed in \cite{fisher1993criticality,levin1996criticality,weiss1998macroscopic,kobrak2008chemical} to study criticality and the gas-liquid transition in Coulomb fluids. 

To determine the fraction of free ions, $\alpha$, we consider a  ``chemical'' equilibrium between free ions and ion pairs, which we define as ions being held at closest approach 
\begin{equation}
\mathrm{cation} \; + \; \mathrm{anion}   \rightleftharpoons \; \mathrm{cation-anion}. 
\end{equation}  
The law of mass action gives the equilibrium constant
\begin{equation}
K = \frac{\rho_d}{\rho_+ \rho_-} =  2 \frac{1-\alpha}{\rho \alpha^2},
\label{mass_action}
\end{equation} 
where $\rho_d$, $\rho_+$ and $\rho_-$ are the densities of dipoles, cations and anions, respectively, and $\rho$ is the total density. Determining $\alpha$ is the central goal of our analysis; ionic liquids would be dilute electrolytes if $\alpha \ll 1$. To relate $K$ to the interaction potential between ions, $v(r)$, we follow the McMillan-Mayer theory of associating liquids \cite{woolley1953representation,levin1996criticality,mcmillan2004statistical}, which yields
\begin{equation}
K = 4 \pi \int_{a}^{\infty}\mathrm{d}r\;  \left[ r^2 (e^{-\beta v(r)} -1)\right]. 
\label{mcmillan_mayer}
\end{equation}

A simple model for charge-charge interaction is the linearised Poisson-Boltzmann approach. Assuming that ions are hard spheres of diameter $a$ that create an exclusion sphere of radius $a$, the interaction potential $v(r)$ between two monovalent ions with opposite charges $\pm e$, can be obtained by solving the linearised Poisson-Boltzmann equation with a uniform charge density on the surface of the exclusion sphere \cite{fisher1993criticality}. This yields
\begin{equation}
\beta v(r) =  -\frac{l_B}{r} \frac{e^{\kappa (a-r)}}{\kappa a +1}. 
\label{PB}
\end{equation}
Equation (\ref{PB}) includes two key lengthscales, the thermal Bjerrum length
\begin{equation}
l_B = \frac{ q^2 e^2}{4 \pi \epsilon \epsilon_0 k_B T},
\label{lb_alpha}
\end{equation} 
which is the typical distance over which electrostatic interactions between two charges in a dielectric medium is comparable to the thermal energy $k_B T$, and the inverse Debye screening length
\begin{equation}
\kappa = \sqrt{4 \pi l_B (\rho_{+} + \rho_-)} =  \sqrt{4 \pi l_B \alpha \rho},
\label{kappa_alpha}
\end{equation} 
which describes the physics of electric field screening by free ions.  

By mass balance and electroneutrality, $\rho_+=\rho_- = \alpha \rho/2$, while the ion pair density $\rho_d = (1-\alpha) \rho/2$. To account for the polarisability of ions, we scale charge and take $q = 1/\sqrt{\epsilon_{\infty}}$ for a monovalent ionic liquid \cite{yu2005accounting,leontyev2009electronic}, where $\epsilon_{\infty}$ is the optical (high frequency) dielectric constant. Such charge-scaling is shown by simulation to give distribution functions that are excellent approximations to the full electrostatic problem with polarisability \cite{schroder2012comparing} . 

Definition (\ref{lb_alpha}) involves $\epsilon$, the static (low frequency) dielectric constant. An ionic liquid is a pure substance and so there is no true ``background'' dielectric. Instead it is the ion pair dipoles that provides the effective dielectric constant. We treat the sea of ion pairs as a dipolar fluid consisting of polarisable spheres with dipole moment $\mu$ whose interactions are screened by free ions. Classic results  \cite{schroer2001generalization} show that in such a system $\epsilon$ satisfies
\begin{equation}
\frac{(\epsilon - \epsilon_{\infty})[(2 \epsilon + \epsilon_{\infty})(1+\kappa a_d) + \epsilon (\kappa a_d)^2 ]}{\epsilon (\epsilon_\infty + 2)^2 (1+\kappa a_d + (\kappa a_d)^2/3) } = \frac{(1-\alpha)  \rho \mu^2 }{18 \epsilon_0 k_B T}, 
\label{diele_DH}
\end{equation} 
where $a_d$ is the effective diameter of the dipole. (Note that in the limit of no free ions, $\kappa = 0$, and (\ref{diele_DH}) reduces to the celebrated Onsager formula \cite{onsager1936electric}.) For simplicity we assume that the ion pair is a sphere with the same volume as the sum of the constituent ions, $i.e.$ $a_d = 2^{1/3} a$ and $\mu = e a_d/2=ea/2^{1/3}$. 
%$\epsilon_{\infty}$ can be calculated by the celebrated Clausius-Mossotti formula \cite{griffiths1999introduction}
%\begin{equation}
%\frac{\epsilon_\infty - 1}{\epsilon_{\infty} + 2} = \frac{1 }{6} (1-\alpha) \rho \nu, 
%\label{clausius}
%\end{equation} 
%with $\nu$ being the polarisability volume and 

Substituting (\ref{mcmillan_mayer}) and (\ref{PB}) into the LHS of (\ref{mass_action}), and using the change of variable $r \rightarrow r/a$, we obtain an implicit equation for $\alpha$, the fraction of ions that are dissociated
\begin{equation}
 \frac{1-\alpha}{\alpha^2} = 2 \pi \rho a^3 \int_{1}^{\infty} r^2 \left[ \exp\left( \frac{l_B}{a r} \frac{e^{\kappa a(1-r)}}{\kappa a +1} \right) -1\right] \; \mathrm{d}r,
\label{eq_for_a}
\end{equation} 
with $\alpha$ implicit in the RHS via (\ref{lb_alpha})-(\ref{diele_DH}), noting that $\epsilon$ is a function of $\alpha$. 

Equation (\ref{eq_for_a}) is the main result of our model and contains no fitting parameters. The only experimental measurements needed are the density $\rho$, ion diameter $a$, and high frequency dielectric constant $\epsilon_{\infty}$. The ion diameter $a$ is difficult to assess experimentally, not least because cations and anions are seldom truly spherical let alone the same size. However, the key dimensionless parameter is $\kappa a$, which can be written in terms of another dimensionless parameter $l_B/a$ via $\kappa a = \sqrt{24 \eta \alpha l_B/a}$, where $\eta = (\pi/6) \rho a^3$ is the packing fraction. It has been shown that $\eta$ is roughly constant for ionic liquids \cite{slattery2007predict}, and we take $\eta = 0.64$ corresponding to random packing of spheres \cite{song2008phase} and comment that the results obtained below are not sensitive to the precise value of the packing fraction used.. It is therefore not necessary to estimate $a$ directly.

The high frequency dielectric constant $\epsilon_{\infty}$ accounts for high-frequency mode for dielectric relaxation \cite{fumino2008cation,weingartner2013static}. Experimentally measured values of $\epsilon_{\infty}$ vary somewhat, see $e.g.$ \cite{daguenet2006dielectric,nakamura2010systematic}, and we will take $\epsilon_{\infty} = 3.5$ as a typical value.

\begin{figure}
\centering
\includegraphics[scale=0.25]{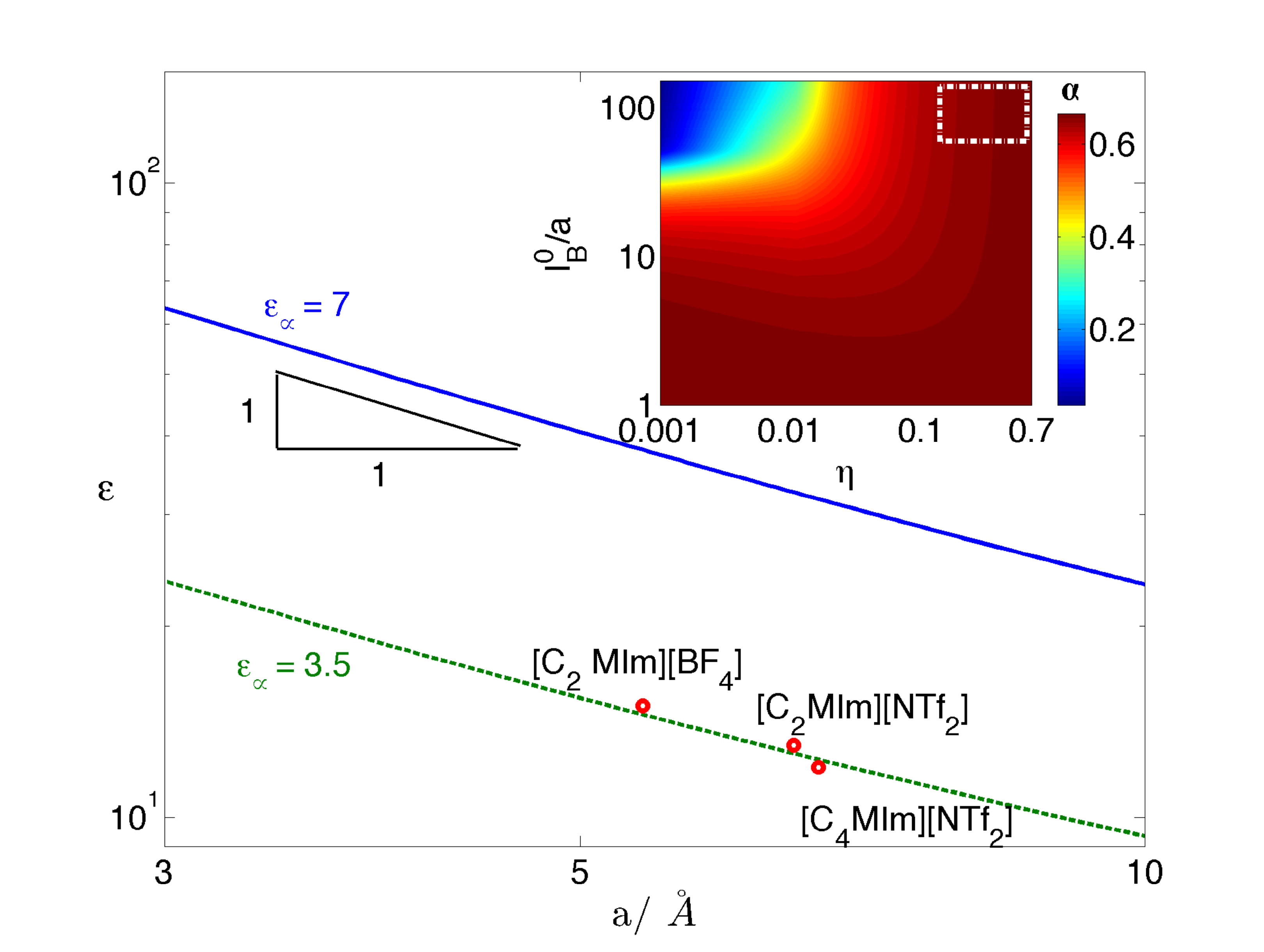}
\caption{The main panel shows the dielectric constant as a function of ion diameter plotted for different high frequency dielectric constants $\epsilon_\infty$. The datapoints show the experimentally obtained values taken from \cite{stoppa2010ideal} ($\mathrm{[C_2 MIm][BF_4]}$), \cite{huang2011static} ($\mathrm{[C_3 MIm][N Tf_2]}$), and \cite{weingartner2006static} ($\mathrm{[C_4 MIm][N Tf_2]}$). The ion diameter is calculated from the density of the ionic liquids taken from \cite{Zhang2009}, assuming the packing fraction $\eta = 0.64$. The inset shows the fraction of free ions, $\alpha$, as a function of packing fraction $\eta$ and unscreened electrostatic interaction at closest contact $l_B^{0}/a$. The white box in the inset shows the largest possible parameter space for a reasonable ionic liquid: ion diameter $5\AA$-$10\AA$ and density $\rho = 4 \times 10^{27} \; \mathrm{m}^{-3} (\approx 6 M)$.  }
\label{diele_plot}
\end{figure}

Figure \ref{diele_plot} shows that the dielectric constants of several common ionic liquids predicted by our simple model quantitatively agree with experimentally measured values. This demonstrates that the thermodynamic framework developed here is physically reasonable. Figure \ref{diele_plot} also shows the general dependence of the dielectric constant on the ionic size $a$. Increasing the effective ionic radius decreases the density of dipoles and hence decreases the dielectric constant. We note that this picture only holds for ionic liquids with short pendant alkyl chains, and is not valid for ionic liquids with longer side chains and polymerised ionic liquids, where locally heterogeneous environments emerge \cite{bhargava2007nanoscale,triolo2007nanoscale,xiao2009effect} \footnote{This promotes ion pairing and correlations between the ion pairs, thus increasing the dielectric constant \cite{choi2013polymerized}.}. Numerical solution of (\ref{eq_for_a}) shows that $\alpha$ is relatively insensitive to the ion radius, but it increases as the polarisability, hence $\epsilon_{\infty}$, increases since Coulomb interactions are better screened (see Fig. \ref{diele_plot}).  

To understand this trend, we note that in the limit $\kappa a \gg 1$, the transcendental equation (\ref{eq_for_a}) can be solved asymptotically, yielding $\alpha \approx 2/3$, which agrees well with numerical results (inset of Fig. \ref{diele_plot}). Physically this result suggests equidistribution between cations, anions and ion pairs, as might be expected from entropy maximisation. The leading order expansion of (\ref{diele_DH}) yields  
\begin{equation}
\epsilon \approx \frac{4 \eta (\epsilon_{\infty} +2)^2}{27} \frac{l_B^{(0)}}{a}, 
\end{equation} 
where $l_B^{(0)}$ is the vacuum Bjerrum length ($\epsilon =1$ in (\ref{lb_alpha})). We see that the dielectric constant scales inversely with the ionic radius, in agreement with numerical results (Fig \ref{diele_plot}). 

Our main results are that $\alpha \approx 2/3$, $i.e.$ almost $2/3$ of ions are \emph{not} bound in ion pairs in a typical ionic liquid at any instant, and that $\kappa a \gg 1$, that is, the Debye length is small, contrary to previous results \cite{gebbie2013ionic}. Although Coulomb interactions in free space are strong and long-ranged, the presence of other free ions significantly screens these interactions.  The inset of Fig \ref{diele_plot} shows that the result $\alpha \approx 2/3$ is robust for the whole parameter space that is physical for ionic liquids. The only parameter regime where ions are strongly bound in ion pairs, and hence $\alpha \ll 1$, is for small ions at low packing fraction. In this regime, the system is best described as a dilute gaseous plasma-like system rather than an ionic liquid.

We should comment on several key assumptions of our model:  $(i)$ The mean-field linearised Poisson-Boltzmann treatment neglects fluctuation effects, which would renormalise the screening length and, for large electrolyte densities, induce an oscillatory decay in the electric potential away from the ion \cite{attard1996electrolytes}. This subtlety is likely to affect the quantitative predictions of our theory, but not our qualitative conclusions as we have accounted for strong electrostatic correlations beyond mean field by considering bound ion pairs. $(ii)$ The estimate of the dielectric constant, (\ref{diele_DH}), uses a mean-field treatment of dipole-dipole interactions, and linearised Poisson-Boltzmann electrostatics (consistent with our treatment of ion-ion interactions). $(iii)$ The calculation of the equilibrium constant is based on McMillan-Mayer theory of an associating solution. The physical picture is that we correct the non-ideality of the ion mixture by introducing dipolar ion-pairs, and Equation (\ref{mcmillan_mayer}) comes from a self-consistent evaluation of the virial coefficients \cite{woolley1953representation}. Various other approaches have been proposed in the literature \cite{levin1996criticality,holovko2005concept,schroer2011chemical}, with various theories differing in their the thermodynamic definition of an ``ion pair''; these yield qualitatively similar results for our model system. $(iv)$ We have only considered ions and ion-pairs here. Interactions between higher order multipoles are weaker. Therefore the fact that even ion-pairs are not abundant means that the effects of larger aggregates would be negligible. $(v)$ Ionic liquids are typically geometrically anisotropic and have short ranged directional interactions such as hydrogen bonds. We have neglected those factors here as they are secondary in the formation of ion pairs.  

%\section{Lifetime of an Ion Pair}
It is important to bear in mind that an ion pair is only a transient species --- ions can dissociate and form an ion pair with another surrounding ion. Therefore, another natural measure of whether ionic liquids are dilute electrolytes is the lifetime of an ion pair. Assuming random packing, the average distance between one ion in an ion pair and another ion of opposite charge is approximately $\sigma =  2[3/(4 \pi \rho)]^{1/3}=a/\eta^{1/3}$. For an ion pair to `break', one ion must move away from the pair. In doing so, the electrostatic energy increases before the ion passes through a ``transition state'' energy maxima, after which it forms a new ion pair. The energy landscape as the ion moves away from its existing partner and towards an adjacent ion is given by
\begin{equation}
V(r) = v(r) + v(\sigma-r+a),
\end{equation}
where $r$ is the separation between the original ion pair (see Figure 3a) and $v(r)$ is given by (\ref{PB}). The lifetime of an ion pair is the mean first passage time through the energy maxima at $r = (\sigma+a)/2$, and can be estimated using Kramers-Smoluchowski theory to be \cite{hanggi1990reaction}
\begin{equation}
\tau = \frac{1}{D} \int_{(\sigma+a)/2}^{\sigma} \mathrm{d}x \; \int_{a}^{x} \mathrm{d} y \; \left(\frac{y}{x} \right)^2 \exp\left[V(x) - V(y) \right], 
\label{eq_kinetics}
\end{equation}   
where $D$ is the self-diffusion coefficient. Table \ref{lifetime} shows that the estimated lifetimes of the ion pairs in typical ionic liquids are relatively short and comparable to the diffusion timescale $\tau_D = a^2/D$. 

The short lifetime, together with the fact that cations, anions and ion pairs exist in roughly equal numbers, suggest that an ion pair is not after all a ``special'' species --- the probability and lifetime of two oppositely charged ions being found within close separation (thus by our definition an ion pair) is almost the same as what one would expect from the relative diffusion of ions in the absence of any interactions.

\begin{table}
    \begin{tabular}{ | l | l | l | l | l |}
    \hline
    Ionic Liquid Ions& $D/10^{8} m^{2} s^{-1}$   & $a/ \AA$ & $\tau$/ps & $p = \tau/\tau_D$  \\ \hline
    $\mathrm{[C_2 MIm][BF_4]}$ & 1.6 \cite{noda2001pulsed} & 5.4 &140 & 7.7 \\ \hline
    $\mathrm{[C_3 MIm][N Tf_2]}$ & 1.2  (est. from \cite{tokuda2005physicochemical})  & 6.5& 71& 2.0 \\ \hline
    $\mathrm{[C_4 MIm][N Tf_2]}$ & 1.2 \cite{tokuda2005physicochemical}   & 6.7&63 & 1.7 \\
    \hline
    \end{tabular}
\caption{The ion pairs have relatively short lifetimes (as estimated by (\ref{eq_kinetics})). $D$ is the diffusion coefficient, $a$ the ion diameter (estimated from the packing fraction, as outlined in the main text), $\tau$ the ion lifetime and $\tau_D = a^2/D$. }
\label{lifetime}
\end{table}

Further insights can be obtained by writing (\ref{eq_kinetics}) in dimensionless form 
\begin{equation}
\tau/\tau_D = p\left(\frac{a}{l_B}, \kappa a\right)
\end{equation}  
where $p$ is a cumbersome quadrature that depends on two dimensionless lengthscales. Fig. 3b shows that, for fixed ion size, $p$ decreases as the dielectric constant, and thus $a/l_B$, increases. However, surprisingly, $p$ increases before decreasing as the Debye length decreases. Decreasing the Debye length decreases the magnitude of the electrostatic interactions, but the energy barrier $\Delta E = V((R+a)/2) - V(a)$ is a non-monotonic function of $\kappa a$. The initial increase in the energy barrier as $\kappa a$ increases is due to the effect of screening decreasing $V(a)$, but to a lesser extent the energy maximum at $V((R+a)/2)$. For larger $\kappa a$ all interactions are strongly screened and therefore $\Delta E$ decreases. Increasing the ion diameter $a$ whilst keeping the Debye and Bjerrum lengths fixed leads to a decrease in the lifetime due to the reduction of surface charge density on the ion surface. 

\begin{figure}\label{energy_well}
\subfigure[]{\includegraphics[scale=0.25]{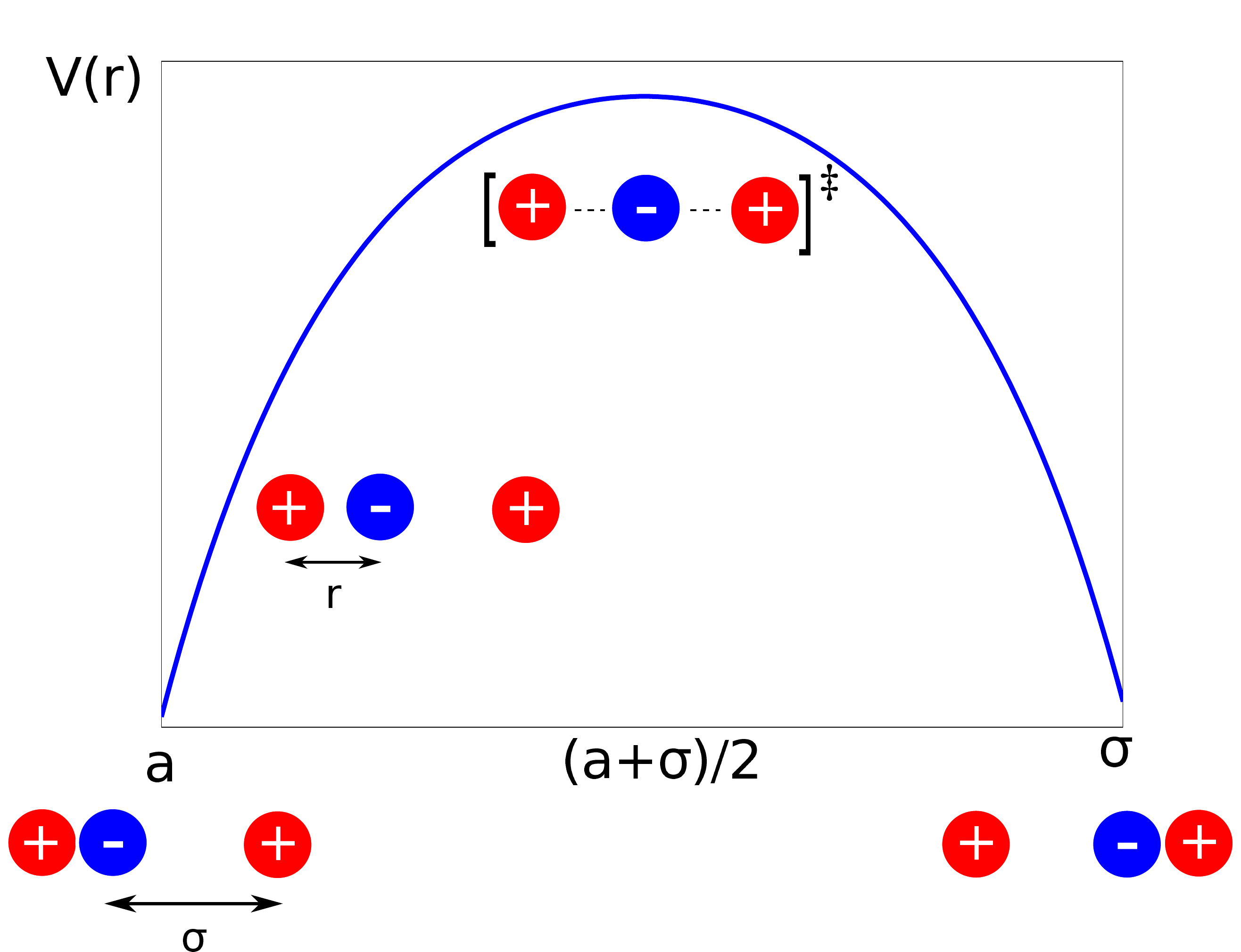}} 
\subfigure[]{\includegraphics[scale=0.2]{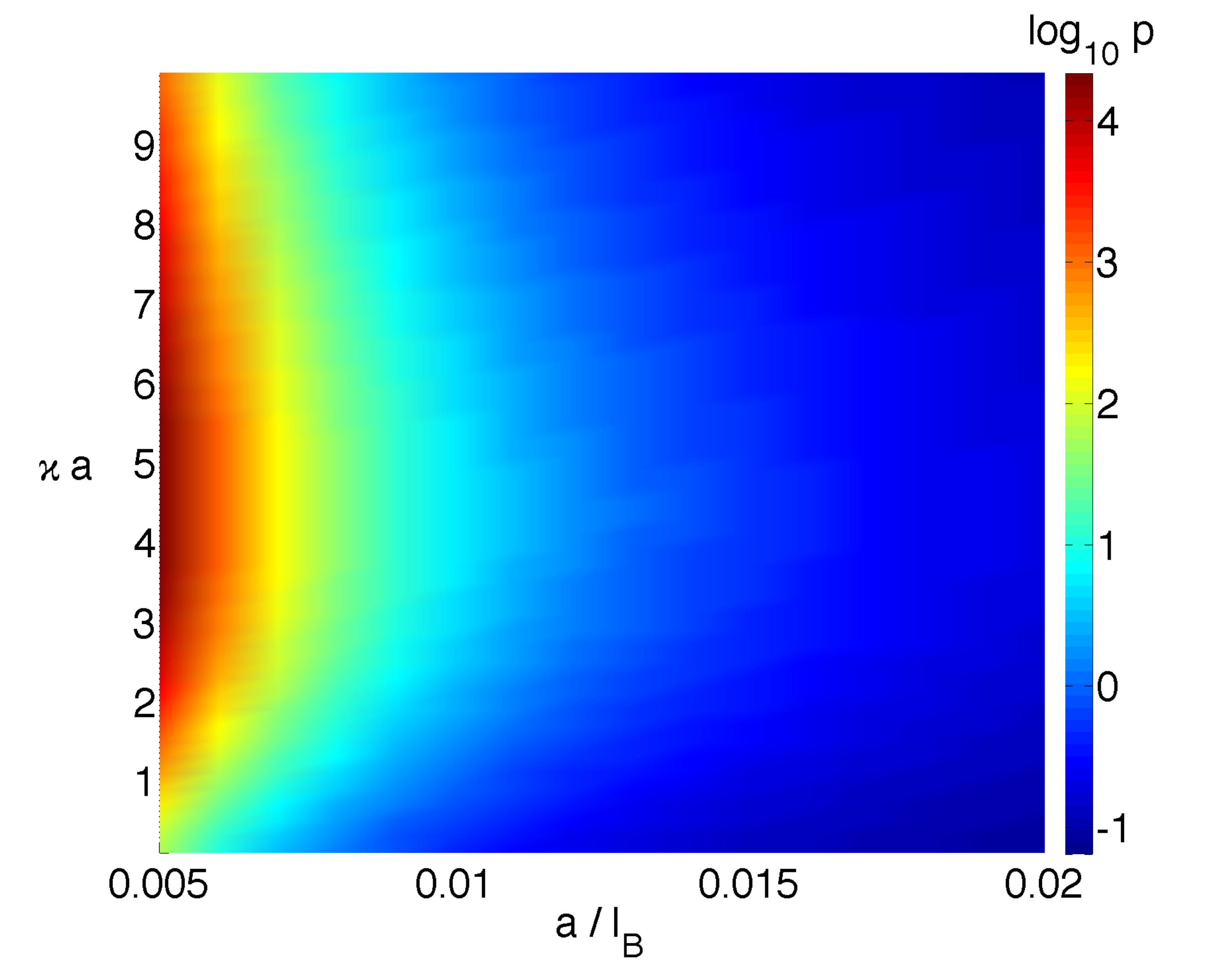}}
\caption{a) Schematic cartoon illustrating the energy landscape of the ion pair exchange. b) The scaling function $p=\tau/\tau_D$ as a function of two dimensionless lengthscales $\kappa a$ and $a/l_B$.}
\end{figure}

%\section{Conclusion}
The simple model developed here captures the essential physics of ion-ion, ion-dipole and dipole-dipole interaction in determining the abundance of ion pairs in ionic liquids. Despite its simplicity, the theory agrees well with available experimental measurements of the dielectric constant. Crucially, our theory predicts that free ions outnumber ion pairs by 2:1 with pairs being short-lived. This prediction suggests that ionic liquids cannot be considered as dilute electrolytes. 

%Simple estimates are obtained for the abundance and lifetime of ion pairs in ionic liquid. It reveals that a significant minority of ions are paired up as short-lived ion pairs. However, the majority of ions exist as free ions, suggesting that ionic liquid cannot be considered to be a dilute electrolyte.  

On the experimental front, we note that the good quantitative fit obtained in \cite{gebbie2013ionic} that supported the dilute electrolyte picture may be an artefact of the surface morphology and interfacial chemistry of gold. It is known that the gold surface is not atomically flat and surface reconstruction occurs upon contact with ionic liquids \cite{aliaga2007surface,uhl2013adsorption}. In fact, other surface force balance studies \cite{perkin2010layering,perkin2011self} seem to indicate a very high degree of Coulomb correlation. We conclude that, unless new  experiments reveal new unexplained behaviour, ionic liquids should not be viewed as dilute electrolytes. 

%We stress that the purpose of this Letter is not to provide a comprehensive thermodynamic model for predicting the physical properties of ionic liquid. Rather, the simple model developed here qualitatively illustrates the importance of adapting a more careful consideration of the roles of free ions and ion pairs in understanding and calculating the association constant and the dielectric constant. 

%We would like to end on a cautionary note. Despite good agreement with measured physical data, w
%The model could be improved by using e.g. integral equation theory instead of linearised Poisson-Boltzmann equation to describe the electrostatics \cite{attard1996electrolytes}, or a more refined definition of association constant \cite{fisher1998exact} to bring the theory into a more quantitative footing. 

\begin{acknowledgments}
We thank D Frenkel for insightful discussions. We also thank the anonymous reviewers for insightful suggestions regarding the interpretation of our results. This work is supported by an EPSRC Research Studentship to AAL.  AG is a Wolfson/Royal Society Merit Award Holder and acknowledges support from a Reintegration Grant under EC Framework VII. 
\end{acknowledgments}

\providecommand{\latin}[1]{#1}
\providecommand*\mcitethebibliography{\thebibliography}
\csname @ifundefined\endcsname{endmcitethebibliography}
  {\let\endmcitethebibliography\endthebibliography}{}

%\bibliography{ref_RTIL}
\end{document}